\documentclass[9pt,twocolumn,twoside]{revtex4}
\usepackage{graphicx}
\usepackage{color}
\usepackage{amsmath}

\begin{document}
\title{Viscoelastic shear stress relaxation in two-dimensional glass forming liquids}
\date{\today}
\author{Elijah Flenner}
\author{Grzegorz Szamel}
\affiliation{Department of Chemistry, Colorado State University, Fort Collins, Colorado 80523, USA}
\begin{abstract}
Translational dynamics of two-dimensional glass forming fluids is strongly influenced by 
soft, long-wavelength fluctuations first recognized by D. Mermin and H. Wagner. 
As a result of these fluctuations, characteristic features of glassy dynamics, such as plateaus 
in the mean squared displacement and the self-intermediate scattering function, are
absent in two dimensions. In contrast, Mermin-Wagner fluctuations do not
influence orientational relaxation and well developed plateaus are observed in
orientational correlation functions. It has been suggested that by monitoring translational 
motion of particles relative to that of their neighbors, one can recover characteristic features
of glassy dynamics and thus disentangle the Mermin-Wagner fluctuations from the 
two-dimensional glass transition. 
Here we use molecular dynamics simulations to study viscoelastic relaxation in
two and three dimensions. We find different behavior of the dynamic modulus below the
onset of slow dynamics (determined by the orientational or cage-relative correlation
functions) in two and three dimensions. The dynamic modulus for two-dimensional supercooled
fluids is more stretched than for three-dimensional supercooled fluids and it does 
not exhibit a plateau, which implies the absence of glassy viscoelastic relaxation.
At lower temperatures, the two-dimensional dynamic modulus 
starts exhibiting an intermediate time plateau and decays similarly to the three-dimensional 
dynamic modulus. 
The differences in the glassy behavior of two- and three-dimensional glass
forming fluids parallel differences in the ordering scenarios in two and three dimensions. 
\end{abstract}

\maketitle

Upon approaching the glass transition, 
a supercooled fluid exhibits
a pronounced viscoelastic response with intermediate time elasticity followed by viscous flow.
This response implies a two-step decay of the viscoelastic response function, \textit{i.e.} of  
the dynamic modulus. 
When a fluid's viscosity reaches $10^{13}$ poise, it appears solid on typical laboratory time 
scales, and this value of the viscosity is often used to 
define the glass transition state point. 
For three dimensional systems, together with pronounced viscoelastic response and 
increasing of a fluid's viscosity, one finds that particle's positions become 
localized for increasingly longer times \cite{Biroli2013,Ediger2012}. 
This localization results in intermediate time plateaus in correlation
functions monitoring particles' translational motion, such as the mean squared displacement
and the intermediate scattering function. 
Flenner and Szamel \cite{Flenner2015} used molecular dynamics simulations to show that
there is no transient localization of particles in two-dimensional glass-forming liquids.
The lack of localization results in an absence of plateaus in the mean squared displacement
and the self-intermediate scattering function. 
This finding has been verified in experiments of quasi-two-dimensional colloidal 
particles \cite{Vivek2017,Illing2017}. 
The absence of transient localization of translational motion originates from 
soft, long wavelength fluctuations \cite{Vivek2017,Illing2017}, which were first 
recognized by D. Mermin and H. Wagner in the context of two-dimensional
magnetic systems \cite{Mermin1966} and crystalline solids \cite{Mermin1968}. 
Importantly, Mermin-Wagner fluctuations 
do not strongly influence orientational relaxation. This is consistent with Flenner and Szamel's
finding of a decoupling between translational and orientational relaxation and the
presence of well developed plateaus in orientational correlation functions in two dimensions. 

It was recently shown that the difference between the two- and three-dimensional glass
transition scenarios evident from Flenner and Szamel's work could be suppressed by
monitoring translational motion of particles with respect to their local environment,
\textit{i.e.} their ``cage'' \cite{Vivek2017,Illing2017,Shiba2018}. This is achieved by 
introducing ``cage-relative'' variants of the mean squared displacement and the intermediate 
scattering function, which use displacements of the particles measured relative to 
their neighbors. These cage-relative functions exhibit well developed 
intermediate time plateaus at state points where the orientational correlation functions 
exhibit plateaus. This observation suggests that to study glassy dynamics of two-dimensional
fluids one needs to remove the effects of Mermin-Wagner fluctuations on the translational
motion by focusing on cage-relative displacements rather than absolute displacements.
The picture that emerged is that when two-dimensional glassy dynamics is viewed in terms 
of local properties, as in bond angle correlations \cite{Flenner2015,Vivek2017},
bond-breaking correlations \cite{Flenner2016,Shiba2012,Shiba2016}, or cage-relative displacements 
\cite{Vivek2017,Illing2017,Shiba2016}, it is quite similar to the three-dimensional 
glassy dynamics \cite{Tarjus2017}.

However, a recent study suggests that two dimensions might play a special role for 
the glass transition and, therefore, glassy dynamics. 
Berthier \textit{et al.}\ \cite{Berthier2018} found that in two dimensions the ideal glass 
transition, defined as the state point at which the configurational entropy vanishes, 
occurs at zero temperature, which is in contrast to the vanishing of the configurational 
entropy at a non-zero temperature in three dimensions \cite{Berthier2017}.

In this work, we address a spectacular manifestation of the incipient glass transition
found in a laboratory: we study the temperature dependence of the viscoelastic response 
and of the shear viscosity. To this end, we monitor the dynamic modulus, which is proportional
to the shear stress autocorrelation function.   
We find that in two dimensions, in the temperature regime where translational
relaxation does not exhibit typical features of glassy dynamics but orientational and
cage-relative translational correlation functions exhibit well developed plateaus, 
the dynamic modulus does not have a plateau. Thus, in this 
temperature regime there is no well developed transient elastic response. At lower 
temperatures, the dynamic modulus develops a plateau signaling emerging viscoelasticity. 
We also find that the shear viscosity grows slower with decreasing 
temperature in two dimensions than in three dimensions, 
and that its growth is decoupled from that of
the orientational and cage-relative relaxation times. 

\section*{Time-dependent viscoelastic response}

\begin{figure}
\includegraphics[width=0.9\columnwidth]{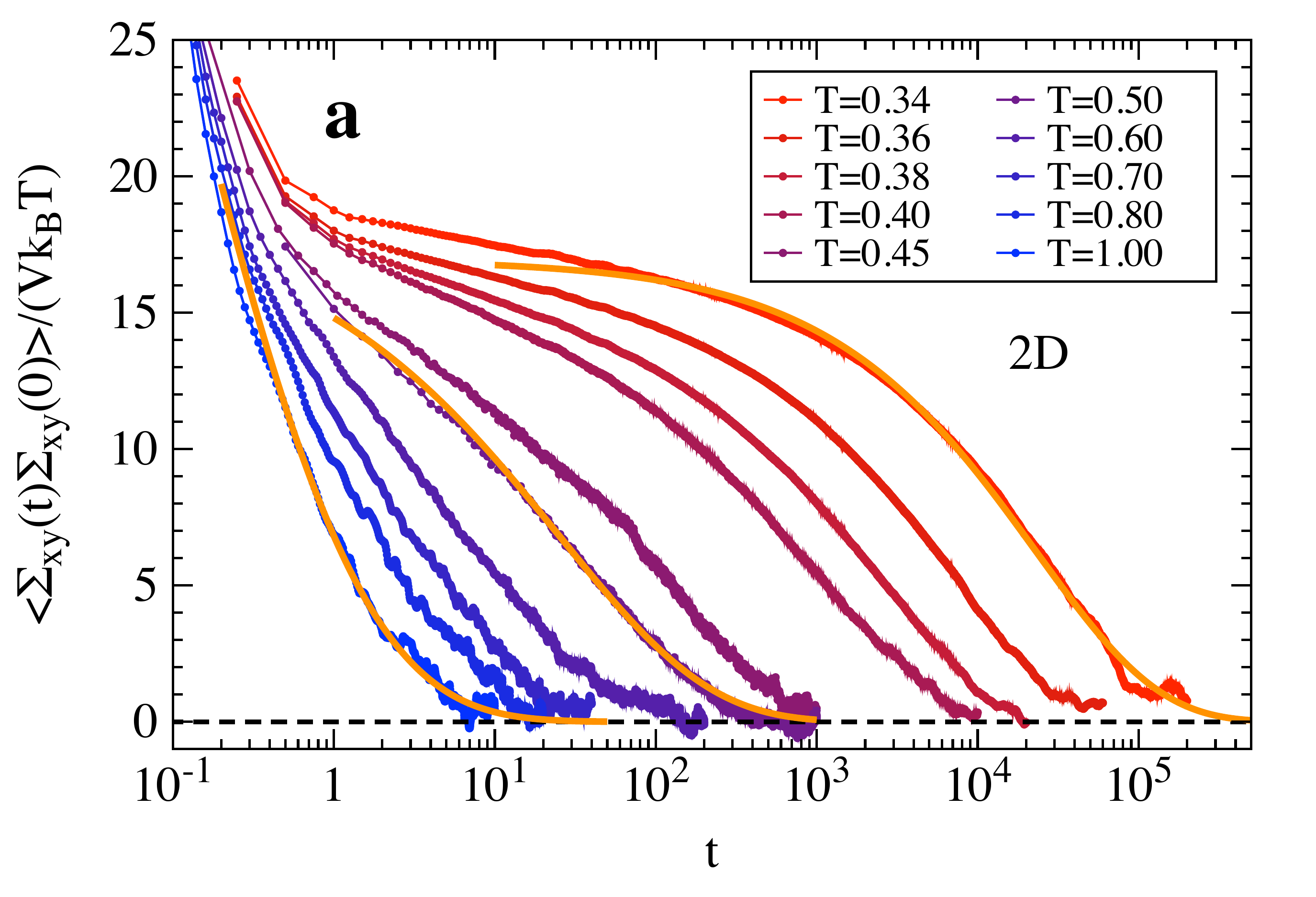}
\includegraphics[width=0.9\columnwidth]{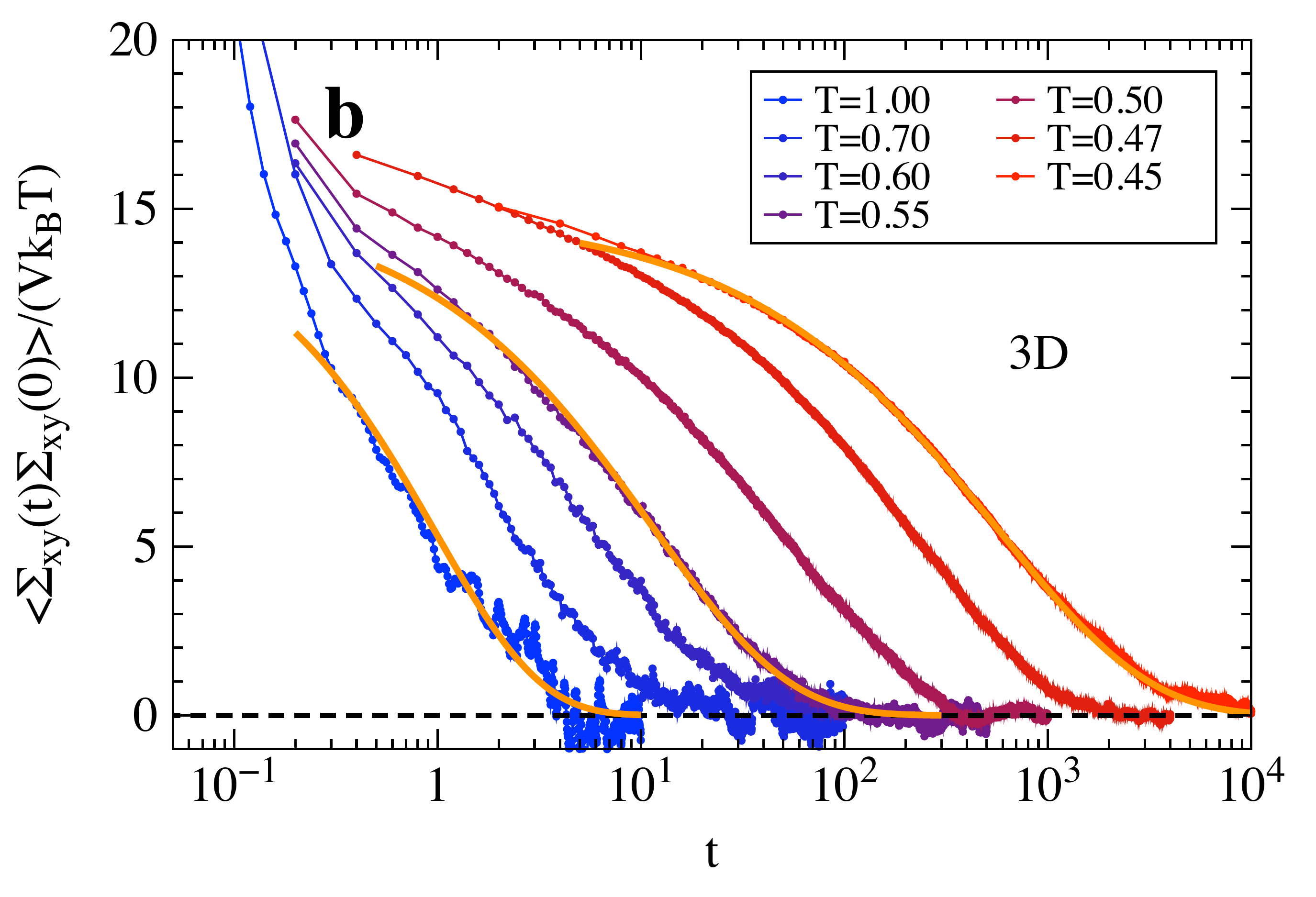}
\caption{\label{fig:ss}Dynamic modulus $G(t)$ for a model two-dimensional glass 
former (a) and a model three-dimensional glass-former (b). The solid orange lines are fits 
to a stretched exponential 
$G_p e^{-(t/\tau)^\beta}$.}
\end{figure} 

The time-dependent viscoelastic response is quantified by the dynamic modulus $G(t)$, which
is proportional to the shear stress autocorrelation function  \cite{EvansMorris}, 
$G(t) = \left< \Sigma_{xy}(t) \Sigma_{xy}(0) \right>/(V k_B T)$,
where $\Sigma_{xy}(t)$ is the $xy$ component of the stress tensor (see Materials and Methods).
We note that the modulus depends on inter-particle distances rather than on absolute
values of particles' coordinates, and thus conceptually it resembles orientational
and cage-relative correlation functions. 

In Fig.~\ref{fig:ss} we compare the dynamic modulus of model two- and three-dimensional
glass-forming systems for temperatures below the onset temperature 
of slow dynamics (as determined by the appearance of intermediate time plateaus in
the orientational and cage-relative correlation functions). 
We checked by simulating systems of different sizes 
that there are no finite size effects for the dynamic modulus in two dimensions, in contrast to
what was observed for the mean squared 
displacement and the self-intermediate scattering function \cite{Flenner2015,Shiba2016}. 
Upon initial visual inspection, at the lowest temperatures accessible in our simulations, 
there appears to be little difference in the time dependence of the dynamic modulus 
for two-dimensional and three-dimensional glass-formers. They both exhibit an initial
decay, an intermediate time plateau, and a final decay from the plateau.
However, a closer examination of Fig.~\ref{fig:ss} and a comparison of the 
temperature dependence of $G(t)$ with that of correlation functions sensitive to
translational, orientational and cage-relative motions reveals important 
differences between two and three dimensions.

\begin{figure}
\includegraphics[width=0.9\columnwidth]{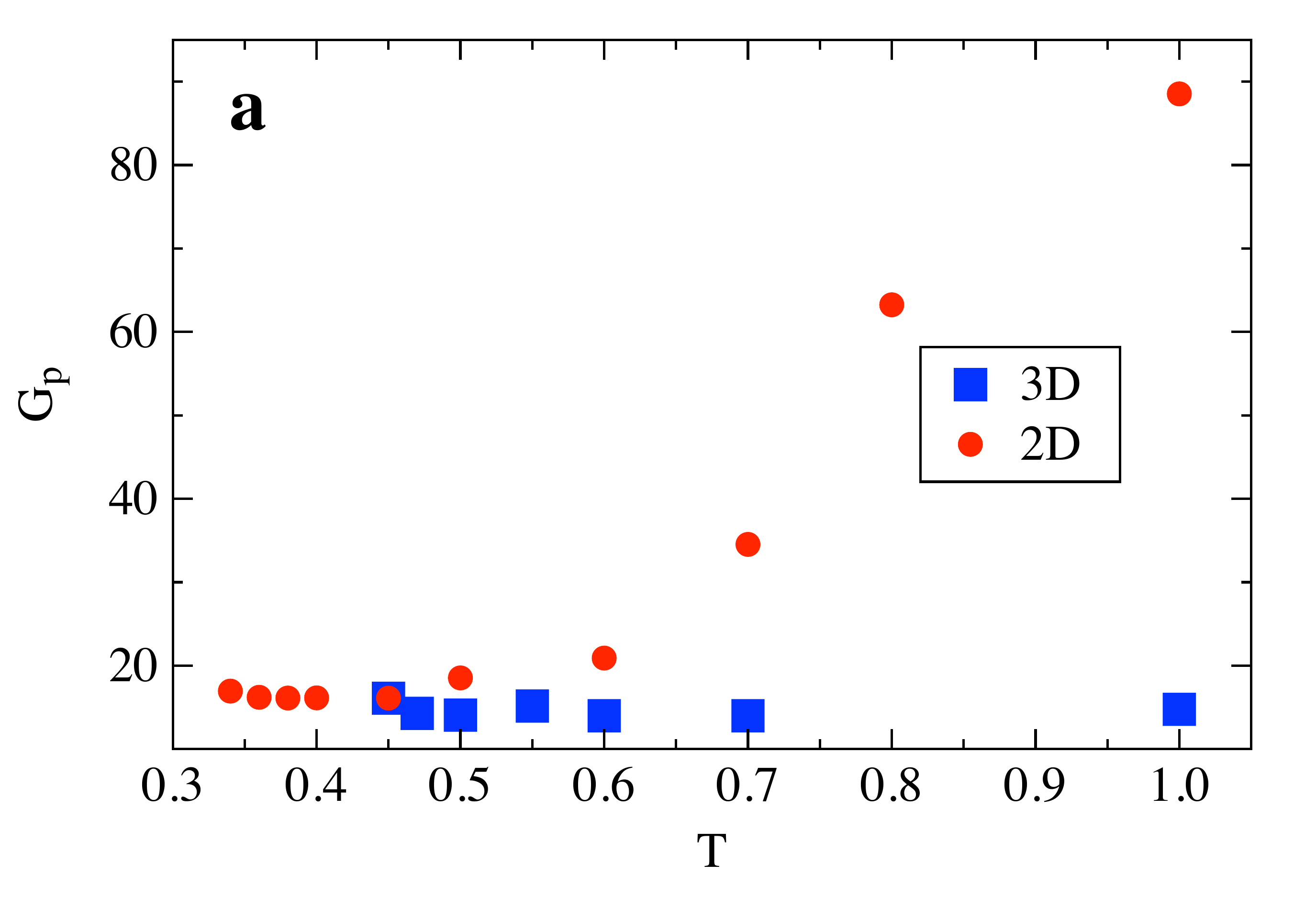}
\includegraphics[width=0.9\columnwidth]{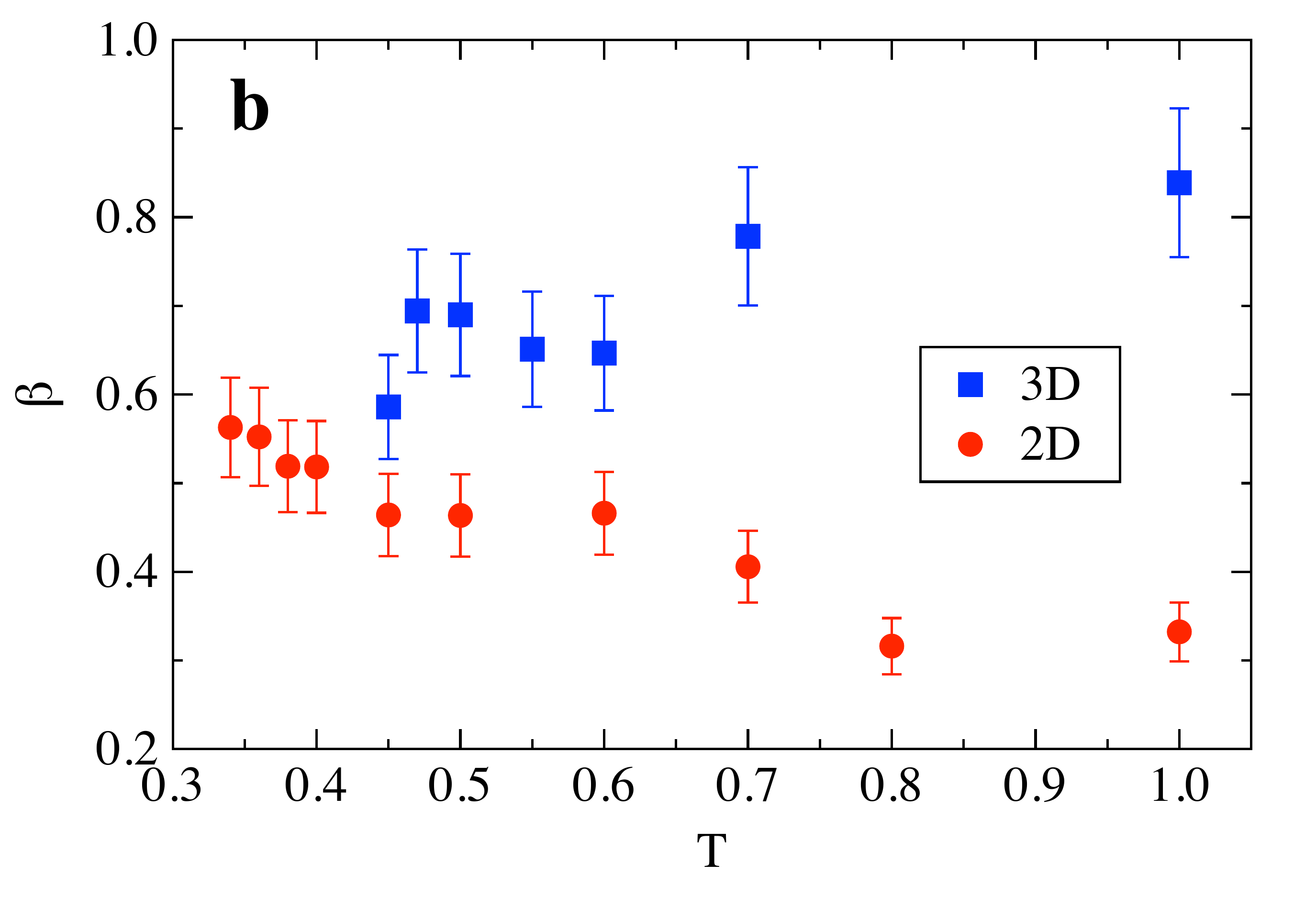}
\caption{\label{fig:Gpbeta}The amplitude of the stretched exponential fit $G_p$ (a) and
the stretching exponent $\beta$ (b) as functions of temperature. 
Just below the onset of slow dynamics there is a
stark contrast between two-dimensional and three-dimensional systems, but 
similar behavior is observed in deeply supercooled fluids.}
\end{figure}

First, we examine the change of the time dependence of the dynamic modulus upon 
decreasing temperature in some detail. 
We fit the final decay to a stretched exponential $G_p e^{-(t/\tau_s)^\beta}$. 
The stretching exponent $\beta$ gives deviations from exponential decay, which is expected for 
three-dimensional liquids well above the onset temperature of 
slow dynamics. The stretched exponential function fits 
the final decay well in every case, and three fits are shown in Fig.~\ref{fig:ss} for
both the two-dimensional and the three-dimensional systems. 
We note that for two-dimensional systems at higher temperatures $G_p$ should only be considered 
a fit parameter; the fit does not does imply existence of an intermediate time plateau in
$G(t)$.

As shown in Fig.~\ref{fig:Gpbeta},
we find very different behavior of the amplitude of the stretched exponential fit $G_p$  
and of the stretching exponent $\beta$ for two-dimensional and 
three-dimensional glass forming systems. 
For two-dimensional glass forming systems at higher temperatures we find that
$G_p$ decreases significantly with decreasing temperature, by a factor of 4 between the onset of
supercooling and the low temperature limit, and $\beta$ increases by a factor of almost 2.
This is in stark contrast to what is observed in three-dimensional systems where 
$G_p$ is approximately temperature-independent and $\beta$ decreases with decreasing temperature. 
These observations and 
a visual inspection of the dynamic modulus shows that 
stress fluctuations relax differently below the onset of supercooling in two- and three-dimensional systems, 
but become similar deep within the supercooled liquid.
This leads to different viscoelastic relaxation; in two-dimensional glass-forming fluids
intermediate-time elastic response appears only upon deep supercooling, whereas in
three-dimensional systems it appears just below the onset of slow dynamics. 

\section*{Shear viscosity}

\begin{figure}
\includegraphics[width=0.9\columnwidth]{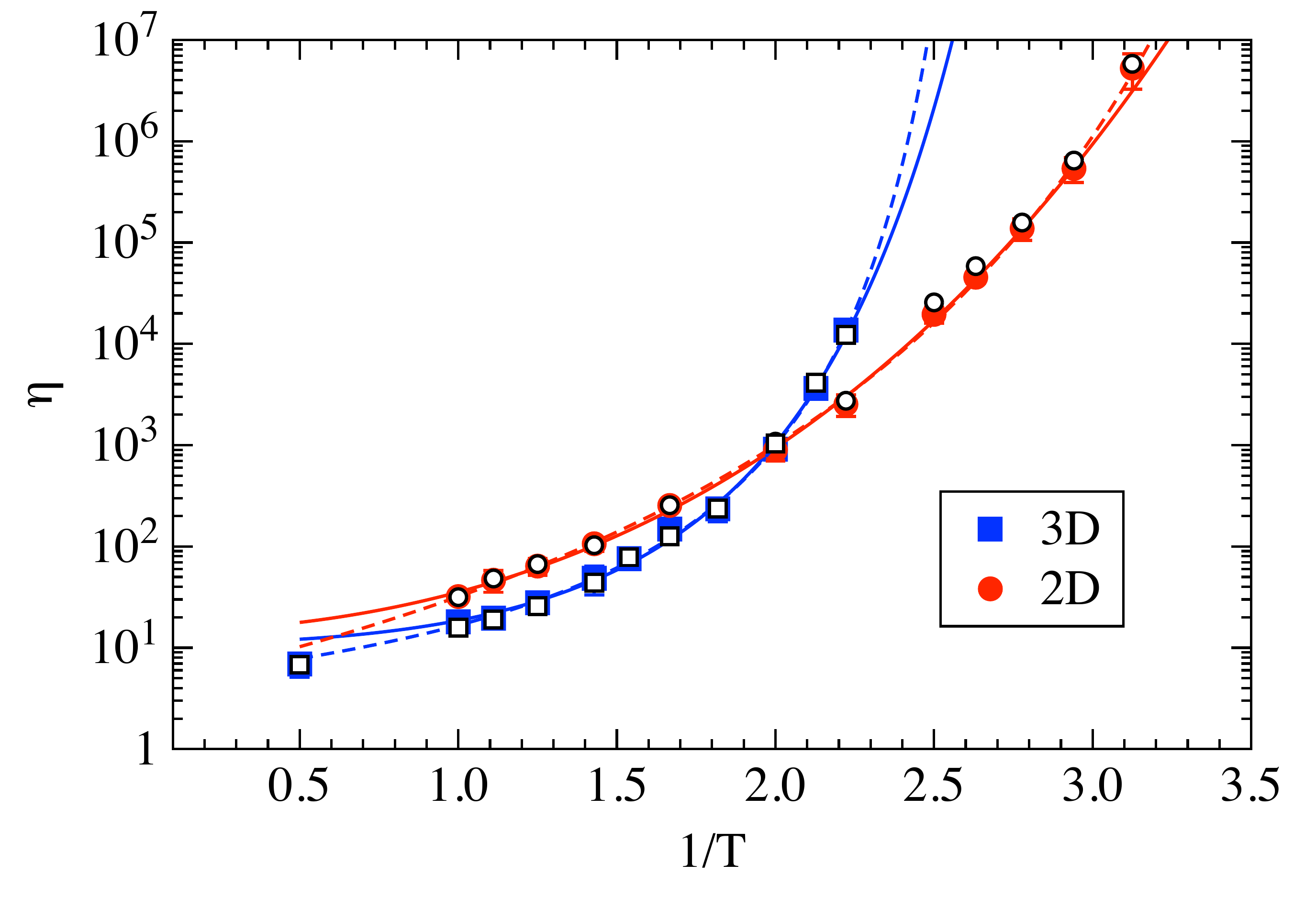}
\caption{\label{fig:visc2d3d}Shear viscosity $\eta$ as a function
of inverse temperature $1/T$. Filled symbols denote $\eta$ calculated from the
exact formula [\ref{etadef}] and open symbols denote $\eta$ as given by
approximate relation [\ref{etaGptau}]. Dashed lines indicate Vogel-Fulcher fits,
$\eta=A\exp(-B/(T-T_0))$, and
solid lines indicate modified Vogel-Fulcher fits, $\eta=C\exp(-D/(T-T_1)^2)$. }
\end{figure}

Dramatic increase of the shear viscosity, $\eta$,
\begin{equation}\label{etadef}
\eta = \int_0^\infty dt\, G(t) \equiv 
\frac{1}{V k_B T}\int_0^\infty dt\, \left< \Sigma_{xy}(t) \Sigma_{xy}(0) \right>
\end{equation}
is a signature of the incipient glass transition.  
In simulations, the range of the increase is limited to 4 or 5 orders of magnitude. 
In Fig. \ref{fig:visc2d3d} we show the shear viscosity for our two-
and three-dimensional glass-forming systems.  
We note that, technically, in two dimensions $\eta$ diverges since there is a long time tail in
$\left<\Sigma_{xy}(t) \Sigma_{xy}(0) \right>$. However, these (hydrodynamic) 
long time tails  are difficult to observe in simulations of glassy systems. In practice 
our upper limit of integration is the first time when our calculated 
$\left<\Sigma_{xy}(t) \Sigma_{xy}(0) \right>$ crosses zero.  
In two dimensions, $\eta$ increases by almost 6 orders of magnitude over the full 
temperature range. 

We observe that the shear viscosity of the two-dimensional system increases rather
gradually with decreasing temperature below the onset of slow dynamics, in contrast to
that of the three-dimensional system. 
We used two different popular fitting functions,
the Vogel-Fulcher fit $\eta=A\exp(-B/(T-T_0))$ and the modified Vogel-Fulcher fit
$\eta=C\exp(-D/(T-T_1)^2)$,
to quantify these temperature dependencies. These fitting functions result in apparent 
glass transition temperatures $T_0$ and $T_1$. Note that while we do not assign any fundamental 
significance to these temperatures, they are useful quantitative characteristics of the 
onset of solid-like response on macroscopic time scales. 

The apparent glass transition temperatures obtained for the two-dimensional glass-former,
$T_0^\text{2d}=0.203\pm 0.010$ and $T_1^\text{2d}=0.068\pm 0.020$, 
are significantly smaller than those
obtained for the three-dimensional glass-former, $T_0^\text{3d}=0.347\pm 0.010$ 
and $T_1^\text{3d}=0.243\pm 0.010$.
This observation, combined with the fact that the onset of slow dynamics 
is the same for both systems, agrees with the observation made in
the previous section. In two dimensions normal features of glassy viscoelastic behavior 
are observed only for deeply supercooled systems. 

In three-dimensional systems, where the intermediate time plateau appears just below the
onset of slow dynamics, viscosity is usually interpreted as a product of the plateau and
the characteristic relaxation time of the final decay. Here we make it more quantitative
and approximate the viscosity by the integral of the final stretched-exponential decay,
\begin{equation}\label{etaGptau}
\eta \approx \int_0^\infty dt\, G_p e^{-(t/\tau_s)^\beta} = G_p \tau_s \Gamma[1+\beta^{-1}],
\end{equation}
where $\Gamma$ denotes the Gamma function. As shown in Fig. \ref{fig:visc2d3d}, for all
temperature below the onset of slow dynamics, approximation
[\ref{etaGptau}] agrees very well with the three-dimensional shear viscosity.

More interestingly, as also shown in Fig. \ref{fig:visc2d3d}, if we apply approximation 
[\ref{etaGptau}] in two dimensions, the result still agrees very well with the 
shear viscosity, for all temperatures below the onset of slow dynamics. This happens in
spite of the fact that for our two-dimensional glass-former, between the onset of slow
dynamics and deep supercooling the dynamic viscosity does not exhibit a clear plateau and the 
fit parameter $G_p$ is strongly temperature dependent.

\section*{Translational, translational cage-relative and orientational 
dynamic correlation functions}

We now compare and contrast temperature dependence of the two-dimensional 
viscoelastic response with that of the two-dimensional translational, 
cage-relative, and orientational time-dependent correlation functions. 

We start by looking at the self-intermediate scattering function 
\begin{equation}
\label{eq:self}
F_s(k;t) = \frac{1}{N}
 \left< \sum_n e^{i \mathbf{k} \cdot (\mathbf{r}_n(t) - \mathbf{r}_n(0))} \right>.
\end{equation}
In Eq.~\ref{eq:self}
$\mathbf{r}_n(t)$ is the position of particle $n$ at a time $t$ 
and $k=|\mathbf{k}|=6.28$ is the peak position of the static structure factor. 
The self-intermediate scattering function
decay time is a measure of the time it takes for a particle is move over
a distance of around $2\pi/k$. For two-dimensional glass-formers, 
due to Mermin-Wagner fluctuations, one would expect $F_s(k;t)$ to decay, albeit possibly 
very slowly. In contrast, for three-dimensional glass-formers intermediate time localization
of the particle positions leads to a pronounced plateau in $F_s(k;t)$.

\begin{figure}
\includegraphics[width=0.9\columnwidth]{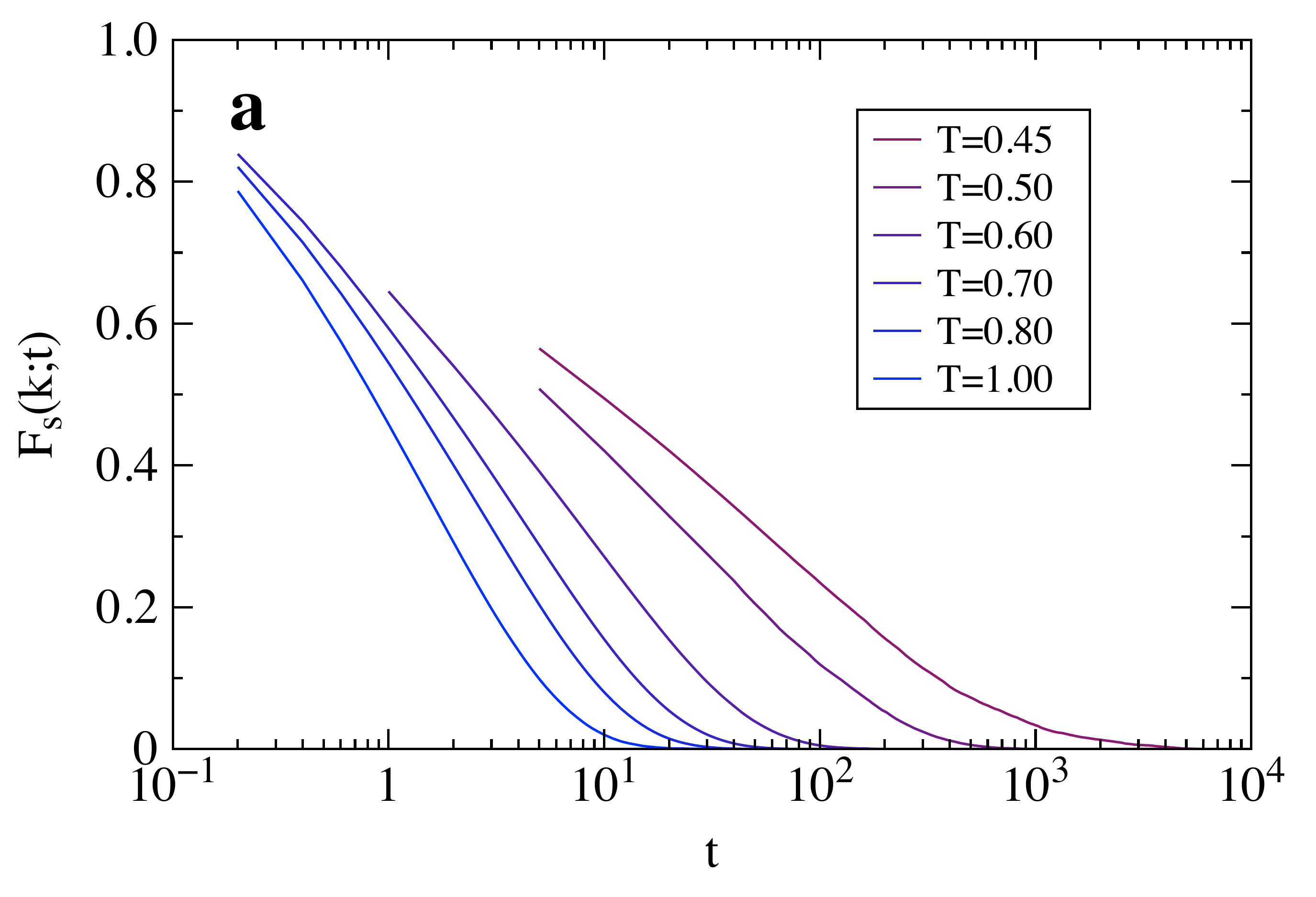}
\includegraphics[width=0.9\columnwidth]{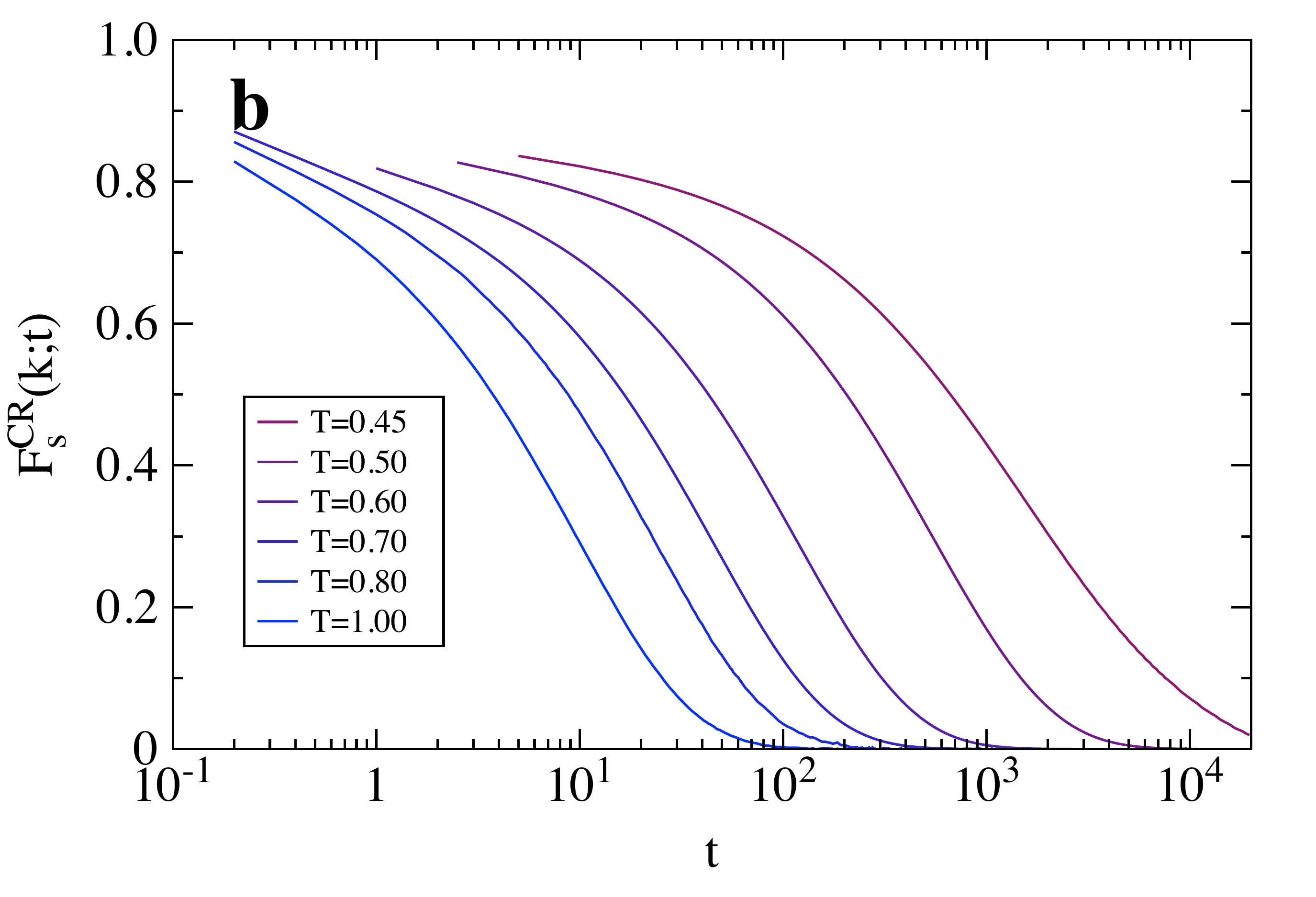}
\caption{\label{fig:fs}a: The self-intermediate scattering function shown for 
the temperature range where we can remove the finite size effects. 
b: The cage-relative self-intermediate scattering function for the same range of temperatures.}
\end{figure}

The self-intermediate scattering function is shown in Fig.~\ref{fig:fs}(a)
for $T=1.0$, 0.8, 0.7, 0.6, 0.5, and 0.45. We note that  
for temperatures below $T=0.45$ we could not simulate systems large enough to 
remove pronounced finite size effects. There is no plateau in the self-intermediate scattering 
function for the temperature range shown in Fig.~\ref{fig:fs}(a).

Recently, simulations and experiments analyzing two-dimensional glass-forming systems have 
examined cage-relative translational dynamics in order to remove the effects of Mermin-Wagner 
fluctuations \cite{Vivek2017,Illing2017,Shiba2018}. 
Examining cage-relative motion is motivated by the study of 
melting of two-dimensional ordered solids. It was argued that by examining cage-relative motion 
one can remove the size effects shown in the dynamics of two-dimensional simulations, restoring
the plateau in the mean squared displacement. Thus, two-dimensional glassy dynamics would then 
have similar characteristics as three-dimensional glassy dynamics. 
A cage-relative self-intermediate scattering function 
\begin{equation}\label{eq:selfcage}
F_s^{CR}(k;t) = \frac{1}{N} \left< \sum_n e^{i \mathbf{k} \cdot \delta \mathbf{r}_n^{CR}(t)}
\right>,
\end{equation}
involves the cage-relative displacements
\begin{equation}
\label{eq:cage}
\delta \mathbf{r}^{CR}_i(t) = [\mathbf{r}_i(t) - \mathbf{r}_i(0)]
- \frac{1}{N_{nn}(i)} \sum_{n=1}^{N_{nn}(i)} [\mathbf{r}_n(t) - \mathbf{r}_n(0)],
\end{equation}
where the summation is over $N_{nn}(i)$ nearest neighbors of particle $i$.
For $k$ around the peak of the static structure factor, $F_s^{CR}(k;t)$ probes the 
escape of a particle from its cage. Shown in Fig.~\ref{fig:fs}b is $F_s^{CR}(k;t)$ for 
temperatures where we could simulate large enough systems to remove finite 
size effects, $T=1.0$, 0.8, 0.7, 0.6, 0.5, and 0.45. We find that the shape of 
$F_s^{CR}(k;t)$ is similar to what is observed for $F_s(k;t)$ for three dimensional 
systems. There is an initial decay to a plateau and then a final decay when the particles begin 
to lose their neighbors. Additionally, we find that time-temperature super-position holds, i.e. 
the final decays of $F_s^{CR}(k;t)$ overlap if scaled by their relaxation time. This is in 
stark contrast to the lack of time-temperature superposition for $F_s(k;t)$. 

However, we recall that there is no plateau 
in the dynamic modulus for temperatures between $T=1.0$ and $T=0.45$.
Thus, by introducing cage-relative function $F_s^{CR}(k;t)$ we
over-emphasize the glassy character of the dynamics.

\begin{figure}
\includegraphics[width=0.9\columnwidth]{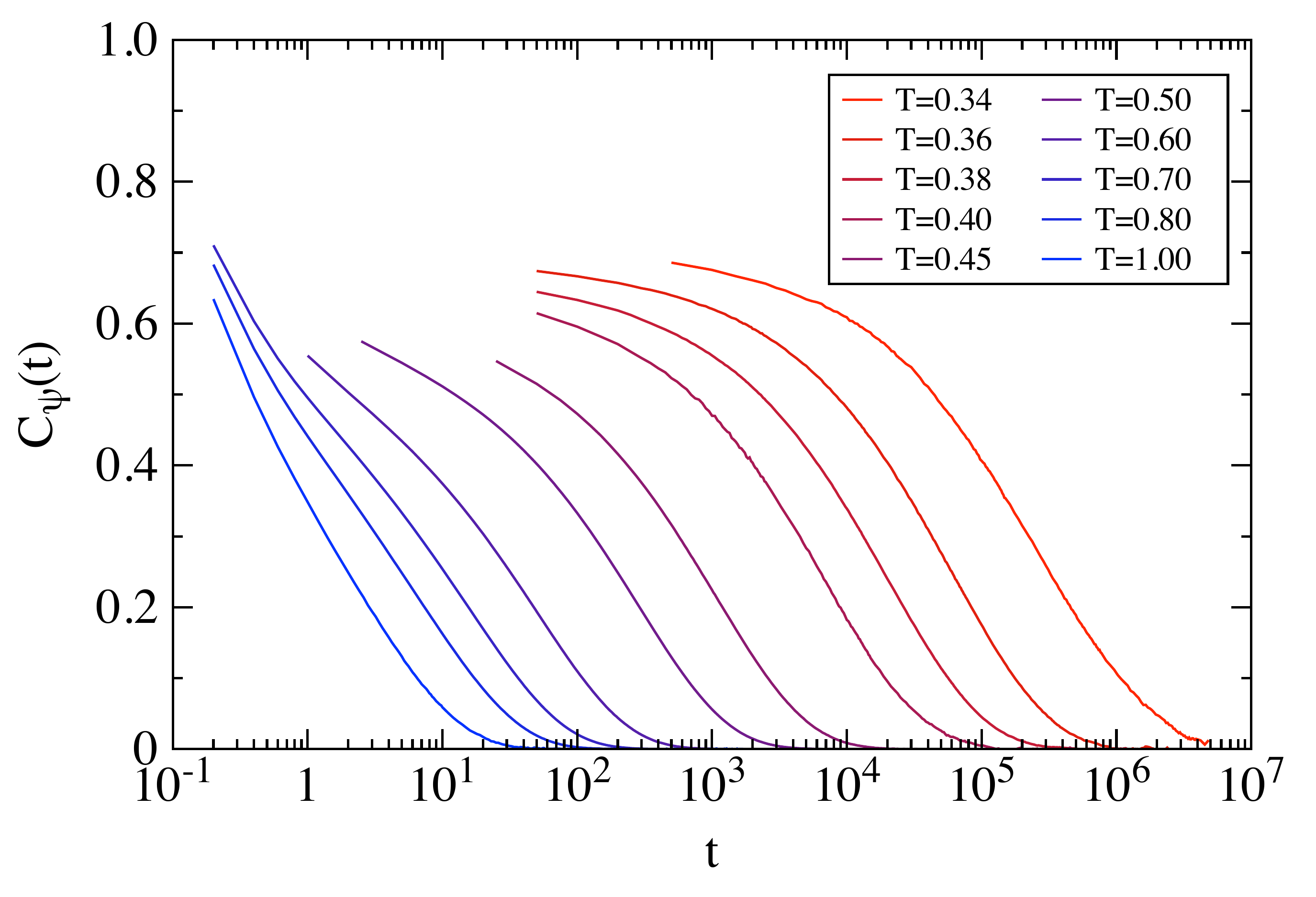}
\caption{\label{fig:psi}The bond angle correlation 
function for the full range of temperatures. The bond angle
correlation function does not exhibit finite size effects.}
\end{figure}
A possible alternative to using cage-relative translational dynamic correlations is provided
by an orientational correlation function 
\begin{equation}
\label{eq:psi}
C_{\psi} = \frac{1}{N} \frac{\left<\sum_n \Psi_6^n(t) \left[\Psi_6^n(0) \right]^*\right>}
{\left< \sum_n \left| \Psi_6^n \right|^2 \right>}.
\end{equation}
In Eq.~\ref{eq:psi} 
\begin{equation}
\Psi_6^n(t) = \frac{1}{N_b} \sum_m e^{i6\theta_{nm}(t)},
\end{equation}
$N_b$ is the number of neighbors identified through Voronoi tessellation, 
$\theta_{nm}(t)$ is the angle that the bond between particle $n$ and particle $m$ makes 
with an arbitrary fixed axis at a time $t$. Shown in Fig.~\ref{fig:psi} is $C_\psi(t)$ 
for our full range of temperatures (we verified that there are no finite size effects
for $C_\psi(t)$). We clearly see the onset of slow dynamics;
at around $T=1$ a plateau emerges. Its height increases slightly with 
decreasing temperature, but the extent of the plateau and the final relaxation time
increase dramatically. 

Again, we emphasize that the orientational correlation function exhibits a plateau 
even when none exists in the dynamic modulus. This is in spite of the fact that 
both functions depend on interparticle distances and, thus, are only sensitive to
local dynamics.

\begin{figure}
\includegraphics[width=0.9\columnwidth]{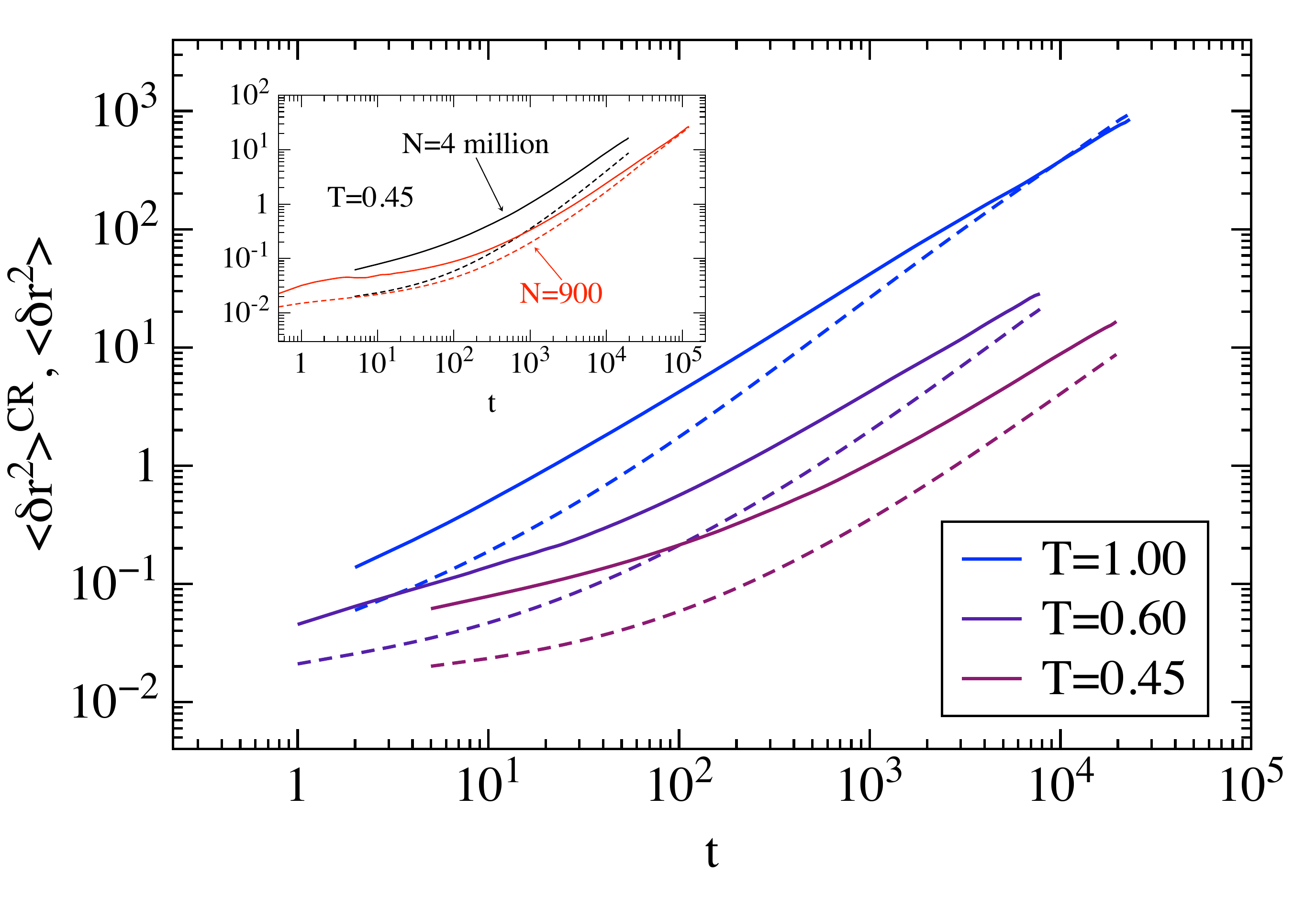}
\caption{\label{fig:crmsd}The cage-relative mean squared displacement 
$\left< \delta r^2(t) \right>^{CR}$ (dashed lines) and the 
mean squared displacement $\left< \delta r^2(t) \right>$ (solid lines) for 
$T=1.0$, 0.60 and 0.45. The inset shows $\left< \delta r^2(t) \right>^{CR}$ and 
$\left< \delta r^2(t) \right>$ 
for $T=0.45$ and two system sizes. There is a pronounced finite size effect at this temperature.}
\end{figure}
Finally, we comment on the expectation that pronounced finite size effects are
removed when one studies cage-relative dynamics. 
For an ordered two-dimensional solid particles stay within a cage formed by their neighbors. 
While the absolute position of the particles drifts due to Mermin-Wagner fluctuations, 
one expects that the position relative to a particles' neighbors remains fixed. 

For a glass in two dimensions, we expect that the cage-relative mean squared displacement
\begin{equation}
\label{eq:msdcage}
\left< \delta r^2(t) \right>^{CR} = 
\frac{1}{N} \sum_i^N \left<\left[ \delta \mathbf{r}_i^{CR}(t)\right]^2\right>,
\end{equation}
will reach a constant value, but 
$\left< \delta r^2(t) \right> = N^{-1} \sum_i \left< [\mathbf{r}_i(t) - \mathbf{r}_i(0)]^2\right>$ 
will drift due to Mermin-Wagner fluctuations. 
In a liquid, for long times the displacements are uncorrelated, 
$\left< [\mathbf{r}_i(t)-\mathbf{r}_i(0)][\mathbf{r}_j(t) - \mathbf{r}_j(0)]\right> 
= \left<[\mathbf{r}_i(t)-\mathbf{r}_i(0)]^2\right>\delta_{ij}$, and then
$\left<\delta r^2(t) \right>^{CR} 
= \left< \delta r^2(t) \right> + \left<N_{nn}\right>^{-1} \left< \delta r^2(t) \right>$. Thus,
in the long-time limit a liquid's cage-relative mean squared displacement is larger than 
the mean squared displacement. 

Shown in Fig.~\ref{fig:crmsd} is $\left< \delta r^2(t) \right>^{CR}$ and  
$\left< \delta r^2(t) \right>$ for $T=1.0$, 0.6 and 0.45. 
There is no plateau in $\left< \delta r^2(t) \right>$, but there is a plateau in 
$\left< \delta r^2(t) \right>^{CR}$.
While the particles displacements are correlated, 
$\left< \delta r^2(t) \right>^{CR} < \left< \delta r^2(t) \right>$. Once the particles 
start to lose their neighbors $\left< \delta r^2(t) \right>^{CR}$ grows faster than 
linearly with $t$ until $\left< \delta r^2(t) \right>^{CR}$
is approximately equal to $\left< \delta r^2(t) \right>$, 
then $\left< \delta r^2(t) \right>^{CR}$ grows linearly with $t$. As shown 
by Shiba \textit{et al.} \cite{Shiba2018} we also find that 
$[\left< \delta r^2(t) \right> - \left< \delta r^2(t) \right>^{CR}] \sim t^{\alpha}$ with 
$\alpha < 1.0$ for a region of time, 
but this difference cannot continue to grow in the liquid and eventually it goes through zero. 

The cage-relative mean squared displacement is also system size dependent in the liquid state. 
Shown in the inset to Fig.~\ref{fig:crmsd} is 
$\left< \delta r^2(t) \right>^{CR}$ and $\left< \delta r^2(t) \right>$
for $N=900$ and $N=4\, 000\, 000$ calculated for $T=0.45$. We can see that 
$\left< \delta r^2(t) \right>^{CR}$ and $\left< \delta r^2(t) \right>$ 
differ at long times. Importantly, there is no finite size effect in the plateau region of  
$\left< \delta r^2(t) \right>^{CR}$. This is in agreement with the observation of 
Shiba \textit{et al.}\ \cite{Shiba2018}. 
These two findings lead us to speculate 
that in the glass the cage-relative mean squared displacement is not system size dependent.

\section*{Time scales}

\begin{figure}
\includegraphics[width=0.9\columnwidth]{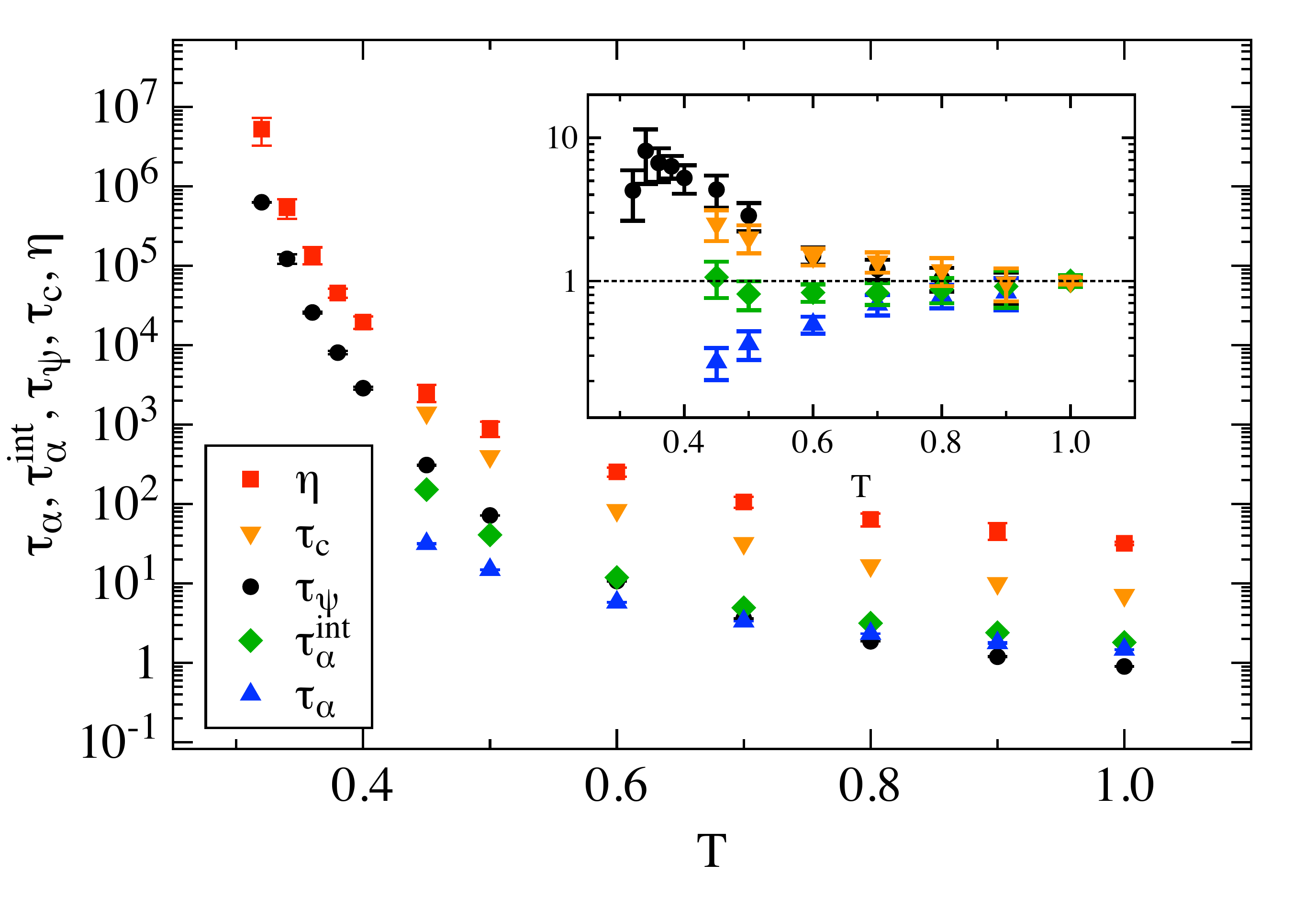}
\caption{\label{fig:tau}The temperature dependence of the relaxation times 
$\tau_\alpha$, $\tau_\alpha^{int}$, $\tau_\psi$, 
$\tau_c$ and the viscosity $\eta$. The inset shows the ratio of the relaxation 
times to the viscosity normalized to this ratio at $T=1.0$.}
\end{figure}

Finally, we briefly compare the growth of the characteristic time scales related to particle 
motion to that of the shear viscosity. We note that the orientational and cage-relative
correlation functions follow time temperature superposition, and thus for these functions we
use the usual definition of the relaxation time,  $C_\psi(\tau_\psi) = e^{-1}$ and
$F_s^{CR}(k;\tau_c) = e^{-1}$. Since $F_s(k;t)$ does not obey time temperature superposition, 
we define two different relaxation times, $F_s(k;\tau_\alpha) = e^{-1}$
and  $\tau_\alpha^{\mathrm{int}} = \int F_s(k;t) dt$.

In Fig.~\ref{fig:tau} we compare the temperature dependence of the relaxation times and 
the shear viscosity. There is only about 2 decades in increase of $\tau_\alpha$ 
(black circles), $\tau_\alpha^{\mathrm{int}}$ (green diamonds), and $\tau_c$ (orange triangles) 
over the temperature range where we can calculate these
relaxation times. However, we do find that 
$\tau_\alpha^{\mathrm{int}}$ increases faster with decreasing temperature than $\tau_\alpha$.
This increase is due to the increasingly stretched behavior of $F_s(k;t)$.  
The orientational relaxation time $\tau_\psi$ is initially slightly smaller than 
$\tau_\alpha$ and $\tau_\alpha^{\mathrm{int}}$, but it increases faster with decreasing 
temperature. We can calculate  $C_\psi(t)$ over the whole temperature and thus
we can see almost 6 orders of magnitude of the orientational relaxation slow down.

To examine the correlations between the relaxation times and the viscosity, we 
calculate the ratios of the relaxation times to the viscosity. 
These ratios, normalized to their values at $T=1.0$, 
are shown in the inset to Fig.~\ref{fig:tau}.  
We find that the cage-relative relaxation time and the orientational relaxation time
increase at statistically the same rate and, importantly, they increase faster than the
shear viscosity. The former result agrees with the observations of 
Refs. \cite{Vivek2017} and \cite{Illing2017}. 
Thus, available evidence suggests that cage-relative and orientational relaxation 
are correlated. However, in the temperature range just below the onset of slow dynamics,
these dynamics do not seem to be correlated with viscoelastic relaxation.

The integrated relaxation time $\tau_\alpha^{\mathrm{int}}$ grows at the same rate as the 
viscosity, while $\tau_\alpha$ defined using the standard definition increases slower than
viscosity with decreasing 
temperature. Therefore, at least in the temperature range in which we can simulate large enough 
systems to remove finite size effects in the self-intermediate scattering function, surprisingly, 
viscoelastic relaxation seems to be correlated with translational dynamics rather than with
orientational relaxation.

\section*{Discussion}

In two dimensions, low temperature phase behavior and properties of ordered solids are 
different from those in three dimensions. In particular, the freezing transition 
can be a two-step process \cite{Strandburg1988,Halperin1978,Nelson1979,Young1979}; 
with semi-long range orientational correlations appearing first and
semi-long range translational correlations and long range (non-decaying) orientational
correlations second. Elasticity appears discontinuously at the second 
transition \cite{Strandburg1988}. 

We showed here that the two-dimensional glass transition scenario reflects the above described
scenario. At the onset of slow dynamics orientational correlations start exhibiting typical 
features of glassy dynamics: two-step decay with intermediate-time plateau whose duration
increases rapidly with decreasing temperature. In contrast, the two-step decay 
is not seen in the self-intermediate scattering function (which is sensitive to translational
relaxation) and in the dynamics modulus (which quantifies viscoelastic relaxation).
At a somewhat lower temperature the latter function (the modulus) starts exhibiting
typical feature of glassy dynamics. Unfortunately, simulating the self-intermediate
scattering function in that temperature range would require using much bigger systems, which
is not possible with our present computational resources. We note that on the basis of
general arguments invoking Mermin-Wagner fluctuations we expect that translational 
motion is not localized even in a very deeply supercooled fluid. If the shear viscosity 
continued to grow as the translational relaxation time, we would
expect that the glass transition to occur at $T=0$. 
We note that this conclusion, and our modified Vogel-Fulcher fit, is compatible with the result 
of Berthier \textit{et al.} \cite{Berthier2018} of a zero-temperature transition.


Finally, we elaborate on a remark we made in passing in the section devoted to
the shear viscosity. In two dimensions, the very existence of a consistent hydrodynamic 
description is in question. Specifically, in two dimensions the so-called hydrodynamic 
long-time tails decay as $1/t$ \cite{Resibois}. 
This suggests that the transport coefficients are divergent.
Somewhat faster decay is obtained if one does a more advanced self-consistent calculation
but the divergence is not eliminated \cite{Resibois,Forster}. 
Here we followed many earlier studies of two-dimensional
glassy phenomena and ignored all these effects. From the practical standpoint, 
the hydrodynamic long-time tails are very difficult to observe in glassy fluids.

\section{acknowledgements}{We gratefully acknowledge partial support of NSF Grant No.~DMR-1608086.
This research utilized the CSU ISTeC Cray HPC System supported by NSF Grant No.~CNS-0923386.}

\section{Simulations}{We simulated the Kob-Andersen binary Lennard Jones mixture in two dimensions and
three dimensions \cite{Kob1994,Kob1995a,Kob1995b}. The interaction potential is 
$V_{\alpha \beta}(r) 
= 4 \epsilon_{\alpha \beta} [(\sigma_{\alpha \beta}/r)^{12} - (\sigma_{\alpha \beta}/)^6]$
where $\epsilon_{BB} = 0.5 \epsilon_{AA}$, $\epsilon_{AB} = 1.5 \epsilon_{AA}$, 
$\sigma_{BB} = 0.8\sigma_{AA}$ and $\sigma_{AB} = 0.88\sigma_{AA}$. The potential is truncated 
and shifted at $2.5\sigma_{\alpha \beta}$. The results are presented in reduced units where 
$\sigma_{AA}$ is the unit of length, $\epsilon_{AA}$ is the unit of energy, and 
$\sqrt{\sigma^2m/\epsilon_{AA}}$ is the unit of time. The mass $m$ is the same for both species. 
For the three dimensional system the larger species composed 80\% of the particles, while the 
larger species composed of 65\% of the particles in the two-dimensional system. We used both 
LAMMPS and HOOMD blue for the two-dimensional simulations, and LAMMPS for the three dimensional 
simulations. We simulated temperatures $T=0.34$, 0.36, 0.38, 0.4, (in 2D) and 0.45, 0.5, 0.6, 
0.7, 0.8, 0.9, and 1.0 (in 2D and 3D). We simulated the system in an NVE ensemble for 
$T \ge 0.5$, but used an NVT Nos\'e-Hoover thermostat with a coupling constant of $\tau=10$. 
All results are averages over four or more production runs. In 2D, for $T \ge 0.9$ we simulated
$10\, 000$ particles. We simulated $250\, 000$ particles for $0.5 \le T \le 0.8$ and 
4 million particles at $T=0.45$. For $T \le 0.4$ we again simulated $10\, 000$ particles. 
We checked to see if there were finite size effects in the shear stress autocorrelation 
function and the bond angle correlation function by simulating $250\, 000$ particles systems 
at $T=0.4$ and comparing them with the $10\, 000$ particle system. The results agreed to 
within error. In 3D, for all temperatures we simulated $27\, 000$ particles.

The $a,b \in (x,y,z)$ component $\Sigma_{a b}$ of the stress tensor is given by 
\begin{equation}
\label{eq:ss}
\Sigma_{ab} = \sum_n^N m_n v_n^a v_n^b - 
\frac{1}{2} \sum_n^N \sum_{m \ne n}^N \frac{r_{nm}^a r_{nm}^b}{r_{nm}} \frac{d V(r_{nm})}{dr_{nm}},
\end{equation}
where $V(r_{nm})$ is the interaction potential between particle $n$ and $m$, $v_n^a$ and 
$r_{nm}^a$ refer to the component of the particle $n$'s velocity and the projection of the 
vector $\mathbf{r}_{nm}$, respectively, along the Cartesian axis $a$. The shear stress is 
the off-diagonal components of $\Sigma_{ab}$. We drop the first term in Eq.~[\ref{eq:ss}] 
since it is small at the temperatures we simulate. The dynamic modulus $G(t)$ is proportional
to the shear stress autocorrelation function \cite{EvansMorris},
\begin{equation}
G(t) = \left< \Sigma_{ab}(t) \Sigma_{ab}(0) \right>/(V k_B T),
\end{equation}
where $a\neq b$, $V$ is the volume of the simulation cell and $k_B$ is Boltzmann's constant. 
}




\end{document}